\documentclass[preprintnumbers,showpacs,amsmath,amssymb,floatfix,9pt,prd,twocolumn,
superscriptaddress,nofootinbib]{revtex4}
\usepackage{latexsym}
\usepackage{epsfig}
\usepackage{amssymb}
\begin{document}

\title{Search of wormholes in different   dimensional   non-commutative
 inspired space-times with Lorentzian distribution  }

\author{Piyali Bhar}
\email{piyalibhar90@gmail.com  } \affiliation{ {  Department of
Mathematics, Jadavpur University, Kolkata 700 032, West Bengal,
India}}

\author{Farook Rahaman}
\email{rahaman@iucaa.ernet.in } \affiliation{ {  Department of
Mathematics, Jadavpur University, Kolkata 700 032, West Bengal,
India}}

\begin{abstract}\noindent
In this paper we are   searching  whether the wormhole
solutions exists in different dimensional noncommutative inspired  spacetimes. It
is well known that the noncommutativity of the space is an outcome
of string theory and it replaced the usual point like object by a
smeared object.   Here we have
chosen  Lorentzian
 distribution as the density function in the noncommutative inspired  spacetime.    We have observed  that the wormhole
 solutions exist  only in four
 and five dimension, however,    higher than fine dimension no wormhole exists. For five dimensional spacetime, we  get a wormhole for a restricted region. In usual four dimensional spacetime, we get a stable wormhole which is asymptotically flat.

\end{abstract}

\maketitle

\section{Introduction}
Wormhole is a hypothetical topological feature of the spacetime
which connects two distinct spacetimes. Morris and
Thorne \cite{morris} have shown that wormhole geometry could be found
by solving Einstein field equation by violating the null energy
condition (NEC) which is threaded as exotic matter. There are a
large number of works
\cite{Lobo,rahaman2,rahaman3} based on the concept of Morris and Throne.\\

 ~In recent years,  researchers have shown considerable interest on
the study of noncommutative spaces. One of the
 most interesting outcomes of the string theory is that the target spacetimes coordinates becomes noncommuting
 operators on D-brane \cite{witten}. Now the noncommutatativity of a spacetime can be encoded in the commutator
  $\left[x^{\mu}, x^{\nu}\right]=i \theta^{\mu \nu}$, where $\theta^{\mu \nu}$ is an anti-symmetric matrix and
  is of dimension $(length)^{2}$ which determines the fundamental cell discretization of spacetime. It is similar
  to the way that the plank constant $\hbar$ discretizes phase space \cite{smailagic}. In noncommutative space the
  usual definition of mass density in the form of Dirac delta function does not hold. So in noncommutative spaces
  the usual form of  the energy density of the static spherically symmetry smeared and
  particlelike gravitational
  source in the form of Lorentzian distribution is
  \cite{Nozari} \[\rho=\frac{M\sqrt{\phi}}{\pi^{2}
  (r^{2}+\phi)^{\frac{n+2}{2}}}\]
Here, the mass M could be the mass of  diffused centralized object such as a wormhole and $\phi$ is noncommutative
 parameter.\\

 Many works inspired by noncommutative geometry are found  in
literature. Nazari{\em et al.}\cite{Nozari}
 used Lorentzian distribution to analyze 'Parikh-Wilczek Tunneling from Noncommutative Higher Dimensional Black Holes'.
 \cite{farook4}  Rahaman {\em et al} have shown that a noncommutative geometrical background is sufficient for
 the existence of a stable circular orbit and one does not need to consider dark matter for galactic rotation curve.
 Kuhfittig\cite{Kuhfittig} found that a special class of thin shell wormholes could be possible that are unstable in classical general
 relativity but   are stable in a small region in noncommutative spacetime. By taking Gaussian distribution as the
  density function Rahaman {\em et al} \cite{farook5} have shown that wormhole solutions exits in usually four as well
   as in five dimensions only. Banerjee \emph{ et al } \cite{banerjee}
    has made a detailed investigation on thermodynamical study e.g.  Hawking temperature, entropy and the area law for
     Schwarzschild black hole in the noncommutative spacetime. Noncommutative Wormholes in $f(R)$ Gravity with
      Lorentzian Distribution has been analyzed in \cite{farook10}. BTZ black hole inspired by noncommutative
      geometry has been discussed in \cite{rahaman10}.\\

 ~Recently, the extension  of general relativity in higher dimension
has become a topic of great interest. The discussion in higher
dimensions is essential due to the fact that many theories indicate
that extra dimensions exist in our Universe. Higher dimensional
gravastar has been discussed by Rahaman et al \cite{fr}.  Rahaman
et al \cite{farook2}   have   investigated whether the
usual solar system tests are compatible with   higher
 dimensions. Another studies in higher dimension are the motion of test particles in the gravitational field of higher dimensional black hole \cite{Liu}.\\

 ~~Inspired by all of these previous work,  we are going to analyze
whether wormhole solutions exists in four and higher
 dimensional spacetime in noncommutative inspired geometry where energy distribution function is taken as  Lorentzian
  distribution. \\

   The plan of our paper as follows: In section II  we have
   formulated basic Einstein field equations. In section III, we have solved those fields equations in different
    dimensions and  in section IV the linearized stability analysis for four dimensional spacetime
    has been worked out. Some discussions and concluding remarks have been done in the final section.\\

\section{Einstein Field Equation in Higher Dimension}

To describe the static spherically symmetry spacetime (in geometrical unit $ G = 1 =c $ here and onwards) in higher
 dimension, we consider  the line element is in the standard form as

\begin{equation}
ds^{2}=-e^{\nu(r)}dt^{2}+e^{\lambda(r)}dr^{2}+r^{2}d\Omega_n^{2}
\end{equation}

where
\begin{equation}
d\Omega_n^{2}=d\theta_1^{2}+\sin^{2}\theta_1d\theta_2^{2}+
\sin^{2}\theta_1\sin^{2}\theta_2d\theta_3^{2}+...+\prod_{i=1}^{n-1}\sin^{2}\theta_id\theta_n^{2}
\end{equation}
The most general form of the  energy momentum tensor for the anisotropic  matter distribution can be taken as \cite{Lobo}
\begin{equation}
T_{\nu}^{\mu}=(\rho+p_r)u^{\mu}u_{\nu}-p_rg^{\mu}_{\nu}+(p_t-p_r)\eta^{\mu}\eta_{\nu}
\end{equation}
 ~~~with $u^{\mu}u_{\mu}=-\eta^{\mu}\eta_{\mu}=1$.

  Here $\rho$ is the energy density, $p_r$ and $p_t$ are respectively the radial and transverse pressures of
 the fluid.\\

In higher dimensions,  Einstein field Equations can be written as \cite{farook5}
\begin{equation}
e^{-\lambda}\left(\frac{n\lambda'}{2r}-\frac{n(n-1)}{2r^{2}}\right)+\frac{n(n-1)}{2r^{2}}=8\pi \rho
\end{equation}
\begin{equation}
e^{-\lambda}\left(\frac{n(n-1)}{2r^{2}}+\frac{n\nu'}{2r}\right)-\frac{n(n-1)}{2r^{2}}=8\pi p_r
\end{equation}

\[\frac{1}{2}e^{-\lambda}\left[\frac{1}{2}(\nu')^{2}+\nu''-\frac{1}{2}\lambda'\nu'
+\frac{(n-1)}{r}(\nu'-\lambda')\right]\]
\begin{equation}
~~~~~~~~~~~~+\frac{(n-1)(n-2)}{2r^{2}}(e^{-\lambda}-1)=8\pi p_t
\end{equation}

Here   ~'prime'~ denotes the differentiation with respect to the radial parameter $r$. \\

In higher dimension,  the density function of a static and spherically symmetric  Lorentzian distribution
of smeared matter as follows \cite{Nozari}
\begin{equation}
\rho=\frac{M\sqrt{\phi}}{\pi^{2}(r^{2}+\phi)^{\frac{n+2}{2}}}
\end{equation}
where $M$ is the smeared  mass distribution.

\section{Model of the Wormhole}

Let us assume the constant redshift function for our model, so called tidal force solution.\\
Therefore,  we have,
\begin{equation}
\nu=\nu_0
\end{equation}
where $\nu_0$ is a constant.\\
Using equation $(8)$ Einstein field equations becomes,
\begin{equation}
\frac{nb'}{2r^{2}}+\frac{n(n-2)b}{2r^{3}}=\frac{8\pi M\sqrt{\phi}}{\pi^{2}(r^{2}+\phi)^{\frac{n+2}{2}}}
\end{equation}
\begin{equation}
8\pi p_r=-\frac{n(n-1)}{2r^{3}}b
\end{equation}
\begin{equation}
8\pi p_t=\frac{(3-n)(n-1)}{2r^{3}}b-\frac{(n-1)b'}{2r^{2}}
\end{equation}
where $b(r)=r(1-e^{-\lambda})$   termed as the shape function
 of the wormhole.\\
From equation $(9)$ we get,
\begin{equation}
b(r)=\frac{16M\sqrt{\phi}}{n\pi}\frac{1}{r^{n-2}}\int\frac{r^{n}}{(r^{2}+\phi)^{\frac{n+2}{2}}}dr+\frac{C}{r^{n-2}}
\end{equation}
Now, the following criterions  need to be imposed for the existence of a wormhole solution.\\
\begin{enumerate}
  \item The redshift function $\nu(r)$ must be finite everywhere ( traversability criteria )
  \item $\frac{b(r)}{r}\rightarrow 0 $ as $n \rightarrow \infty $ ( asymptotically flat spacetime )
  \item At the throat $(r_0)$ of the wormhole $b(r_0)=r_0$ and $b'(r_0)<1$ ( flaring out condition )
  \item $ b(r)<r $ for $r>r_0$  (regularity of metric coefficient)

\end{enumerate}

\subsection{Four dimensional Spacetime (n=2)}
In four dimensional spacetime,
the shape function of the wormhole is given by,
\begin{equation}
b(r)=\frac{4M\sqrt{\phi}}{\pi}\left\{\frac{1}{\sqrt{\phi}}\tan^{-1}(\frac{r}{\sqrt{\phi}})-\frac{r}{r^{2}+\phi}
 \right\}+C
\end{equation}
where, C is an integration constant.

The radial and transverse pressure can be obtained as
\begin{equation}
8\pi p_r=-\frac{1}{r^{3}}\left[\frac{4M\sqrt{\phi}}{\pi}\left\{\frac{1}{\sqrt{\phi}}\tan^{-1}(\frac{r}{\sqrt{\phi}})
-\frac{r}{r^{2}+\phi}\right\}+C\right]
\end{equation}

\pagebreak

\[8\pi p_t=\frac{1}{2r^{3}}\left[\frac{4M\sqrt{\phi}}{\pi}\left\{\frac{1}{\sqrt{\phi}}\tan^{-1}(\frac{r}{\sqrt{\phi}})
-\frac{r}{r^{2}+\phi}   \right\}+C \right]\]
\begin{equation}
 ~~~-\frac{4M\sqrt{\phi}}{\pi}\frac{1}{(r^{2}+\phi)^{2}}
\end{equation}


\begin{figure}[htbp]
    \centering
        \includegraphics[scale=.3]{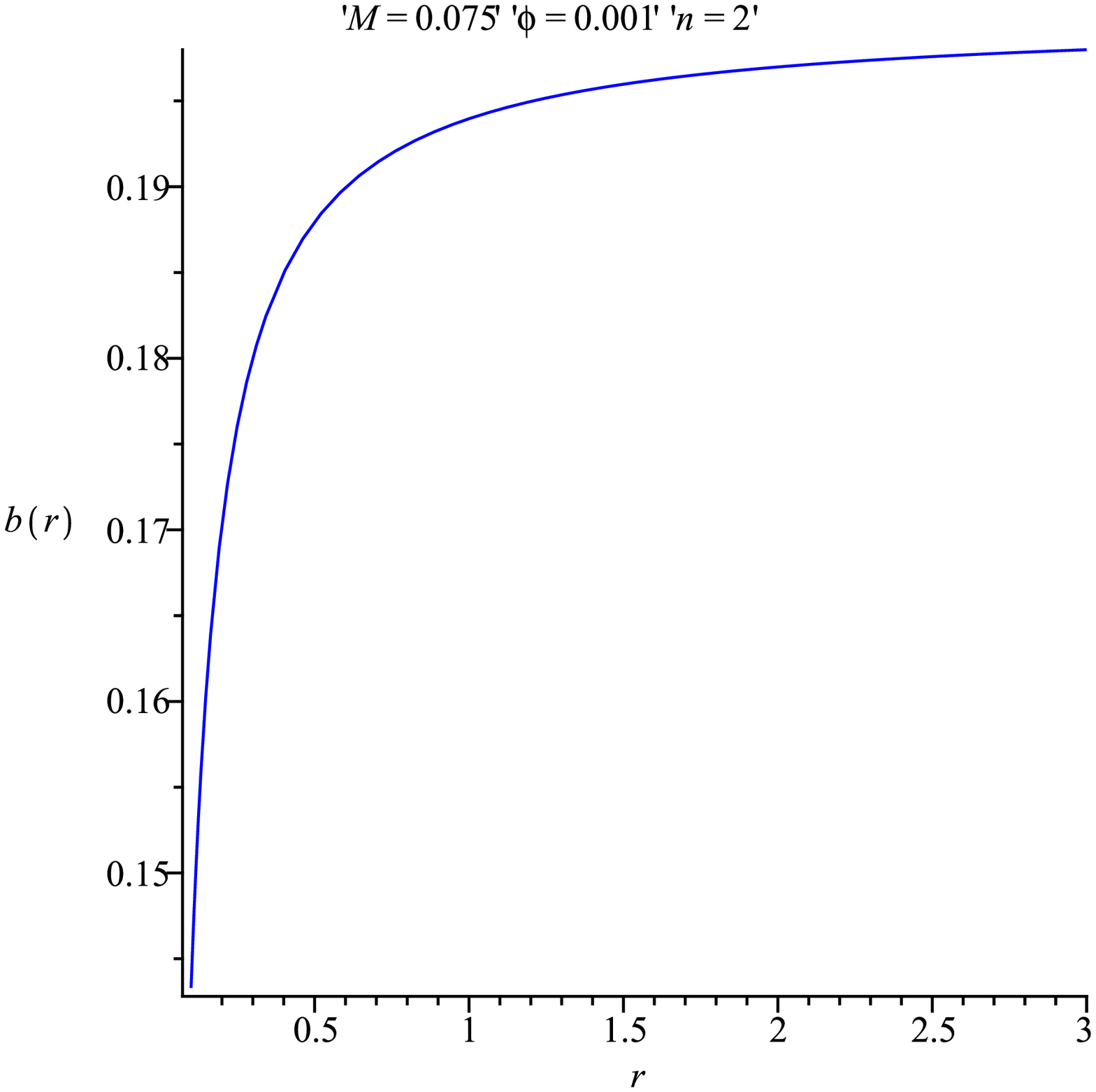}
       \caption{ Diagram of the shape function of the wormhole in four dimension is shown against $r$.}
    \label{fig:1}
\end{figure}

\begin{figure}[htbp]
    \centering
        \includegraphics[scale=.3]{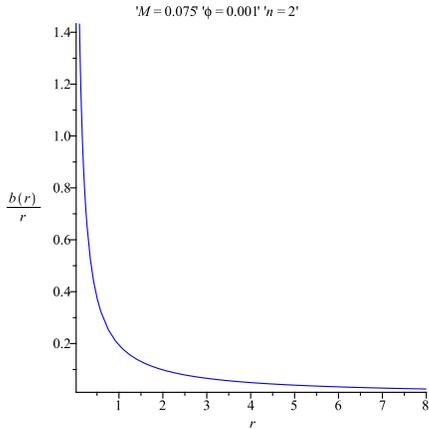}
       \caption{Asymptotically flatness condition in four dimensional spacetime is shown against $r$.}
    \label{fig:1}
\end{figure}

\begin{figure}[htbp]
    \centering
        \includegraphics[scale=.3]{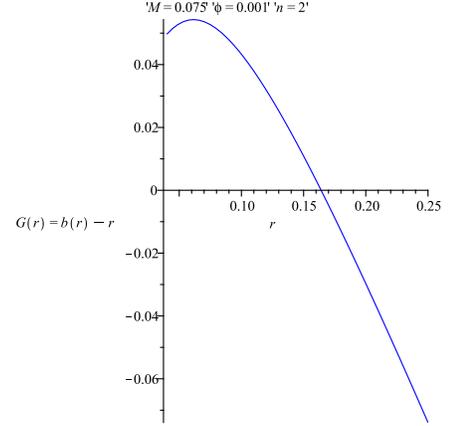}
       \caption{The throat of the wormhole in $4D$ occurs where G(r)cuts the 'r'axis.}
    \label{fig:1}
\end{figure}

\begin{figure}[htbp]
    \centering
        \includegraphics[scale=.3]{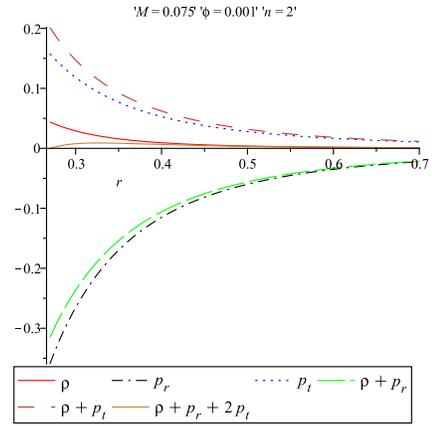}
       \caption{ The energy condition in $4D$ is shown against $r$}
    \label{fig:1}
\end{figure}

Next, we are going to verify whether the shape function $b(r) (see~ FIG.~1)$ satisfies all the physical requirements
 to form a wormhole structure (which has been given in section III). For this purpose we assign some arbitrary values
 to the parameters ( mentioned in the figures). Form $FIG.~2$, we see that $\frac{b(r)}{r}\rightarrow 0 $ as $r~\rightarrow ~\infty $ which verifies
  that the spacetime is asymptotically flat. The throat $r_0$ of the wormhole occurs where $b(r_0)=r_0$,  which is the
  root of the function $G(r)=b(r)-r$. We have plotted $G(r)~Vs.~r$ in $FIG.~3$. The throat of the wormhole occurs
   where $G(r)$ cuts the $r$ axis.   From this figure, we see that $r_0=0.165$. $FIG.~3$ also indicates that $G(r)<0$
   when $r > r_0$ which implies $\frac{b(r)}{r}<1$ when $r > r_0$. One can also notice from equation $(9)$ that at
   the throat of the wormhole $(r_0=0.165)$ $b'(0.165)=0.207 < 1$. Hence Flare-out condition  is also satisfied.

According to Morris \& throne \cite{morris} the embedding surface of the wormhole is denoted by the function $z(r)$
which satisfies the following differential equation
\begin{equation}
\frac{dz}{dr}=\pm\frac{1}{\sqrt{\frac{r}{b(r)}-1}}
\end{equation}
The above equation $\frac{dz}{dr}$ diverges at the throat of the
 wormhole which concludes   that the embedding surface is vertical at the throat.\\
Equation $(16)$ gives,
\begin{equation}
z=\pm\int_{r_0^{+}}^{r}\frac{dr}{\sqrt{\frac{r}{b(r)}-1}}
\end{equation}
The proper radial distance of the wormhole is given by
\begin{equation}
l(r)=\pm\int_{r_0^{+}}^{r}\frac{dr}{\sqrt{1-\frac{b(r)}{r}}}
\end{equation}
(here, $r_0$ is the throat of the wormhole.)
The above two integrals mentioned in equation (17) and (18) can not be done analytically. But we can perform it numerically. The numerical values are
obtained by fixing a particular value of lower limit and by changing the upper limit, which are given in Table-I
 and Table-II respectively.
The embedding diagram $z(r)$ and the radial proper distance $l(r)$ of 4D wormhole have been depicted in fig.~5 and fig.~7 respectively.Fig.~6 represents the full visualization of 4D wormhole obtained by rotating embedding curve about 'z-axis'.

We can match our interior wormhole spacetime with the Schwarzschild exterior spacetime
\begin{equation}
ds^{2}=-\left(1-\frac{2m}{r}\right)dt^{2}+\frac{dr^{2}}{1-\frac{2m}{r}}+r^{2}d\Omega_2^{2}
\end{equation}
where \[d\Omega_2^{2}=d\theta^{2}+\sin^{2}\theta d\phi^{2}\]

\begin{table}
\caption{Values of $z(r)$  for different $r$. $ r_0^{+} =0.17$,
$\phi=0.001$,$M=0.075$}
{\begin{tabular}{@{}cc@{}} \toprule $r$ & $z(r)$ \\
\colrule
2  &  ~~~~0.9462904189\\
4  &  ~~~~1.476685471\\
6  &  ~~~~1.881582603\\
8  &  ~~~~2.222220748\\
10 &  ~~~~2.521990341\\
12 &  ~~~~2.792809650\\
\botrule
\end{tabular}}
\end{table}

\begin{table}
\caption{Values of $l(r)$  for different $r$. $ r_0^{+} =0.17 $,
$\phi=0.001$,$M=0.075$}

{\begin{tabular}{@{}cc@{}} \toprule $r$ & $l(r)$ \\
\colrule
2  &  ~~~~2.053236413\\
4  &  ~~~~4.106613071\\
6  &  ~~~~6.137481048\\
8  &  ~~~~8.159285247\\
10 &  ~~~~10.17615752\\
12 &  ~~~~12.18992251\\
14 &  ~~~~14.20154866\\

\botrule
\end{tabular}}
\end{table}
Now by using Darmois-Israel\cite{isreal1,isreal2} junction condition the surface energy density ($\sigma$) and surface
pressure ($\mathcal{P}$) at the junction surface $r=R$ can be obtained as,
\begin{equation}
\sigma=-\frac{1}{4\pi R}\left[\sqrt{1-\frac{2m}{R}+\dot{R}^{2}}-\sqrt{1-\frac{b(R)}{R}+\dot{R}^{2}}\right]
\end{equation}
\begin{widetext}
\begin{equation}
\mathcal{P}=\frac{1}{8\pi R}\left[\frac{1-\frac{m}{R}+\dot{R}+R\ddot{R}}{\sqrt{1-\frac{2m}{R}+\dot{R}^{2}}}-
\frac{1-\frac{b(R)}{R}
+\dot{R}^{2}+R\ddot{R}-\frac{\dot{R}^{2}}{2}\frac{b-b'R}{R-b}}{\sqrt{1-\frac{b(R)}{R}+\dot{R}^{2}}}\right]
\end{equation}
\end{widetext}

The mass of the thin shell $(m_s)$ is given by
\begin{equation}
m_s=4\pi R^{2}\sigma
\end{equation}
Using the expression of $\sigma$ in equation $(20)$ (considering the static case) one can obtain the mass of the
 wormhole as
\begin{equation}
m=\frac{b(R)}{2}-\frac{m_s}{2}\left[\frac{m_s}{R}+2\sqrt{1-\frac{b(R)}{R}}\right]
\end{equation}
This gives the mass of the wormhole in terms of the thin shell mass.\\

~~~~Let us use the conservation identity  given by $$S_{j,i}^{i}=-[\dot{\sigma}+2\frac{\dot{R}}{R}(\mathcal{P}
+\sigma)]$$ which gives the following expression
\begin{equation}
\sigma'=-\frac{2}{R}(\mathcal{P}+\sigma)+\Xi
\end{equation}
where $\Xi$ is given by,
\begin{equation}
\Xi=-\frac{1}{4\pi R}\frac{b-b'R}{2(R-b)}\sqrt{1-\frac{b(R)}{R}+\dot{R}^{2}}
\end{equation}
The above expression $\Xi$ will be used to discuss the stability analysis of the wormhole.\\
  From equation (22) we get,
\[\left(\frac{m_s}{2a}\right)''=\Upsilon-4\pi \sigma'\eta\]
where \[\eta=\frac{\mathcal{P}'}{\sigma'}~~~~~~~~~and~~~\Upsilon=\frac{4\pi}{R}(\sigma+\mathcal{P})+2\pi R\Xi'\]
where the parameter $\sqrt{\eta}$ is generally denotes the velocity of the sound. So one must have $0<\eta\le1$ .

\begin{figure}[htbp]
    \centering
        \includegraphics[scale=.3]{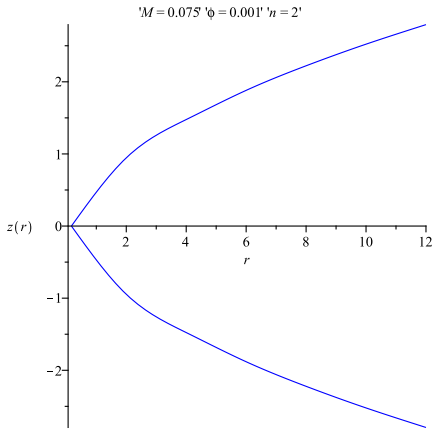}
       \caption{The embedding diagram of 4D wormhole has been shown against $r$}
    \label{fig:1}
\end{figure}

\begin{figure}[htbp]
    \centering
        \includegraphics[scale=.3]{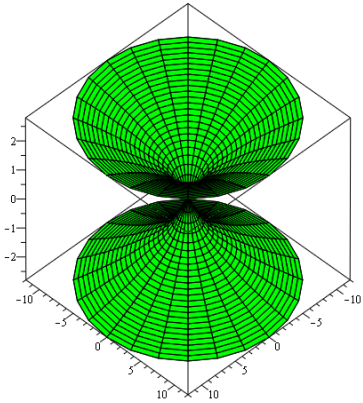}
       \caption{The full visualization of 4D wormhole surface is obtained by rotating the embedding curve  with respect to 'z-axis'}
    \label{fig:1}
\end{figure}

\subsection{Five dimensional Spacetime (n=3)}
In five dimensional spacetime,
the shape function of the wormhole is given by,
\begin{equation}
b(r)=\frac{1}{r}\left\{-\frac{16M\sqrt{\phi}}{9\pi}\frac{3r^{2}+2\phi}{(r^{2}+\phi)^{\frac{3}{2}}}+C\right\}
\end{equation}
The radial and transverse pressures can be obtained as,
\begin{equation}
8\pi p_r=-\frac{3}{r^{4}}\left\{-\frac{16M\sqrt{\phi}}{9\pi}\frac{3r^{2}
+2\phi}{(r^{2}+\phi)^{\frac{3}{2}}}+C\right\}
\end{equation}

\begin{equation}
8\pi p_t=\frac{1}{r^{4}}\left\{-\frac{16M\sqrt{\phi}}{9\pi}\frac{3r^{2}
+2\phi}{(r^{2}+\phi)^{\frac{3}{2}}}+C\right\}-\frac{16M\sqrt{\phi}}{3\pi}\frac{1}{(r^{2}+\phi)^{\frac{5}{2}}}
\end{equation}

\begin{figure}[htbp]
    \centering
        \includegraphics[scale=.3]{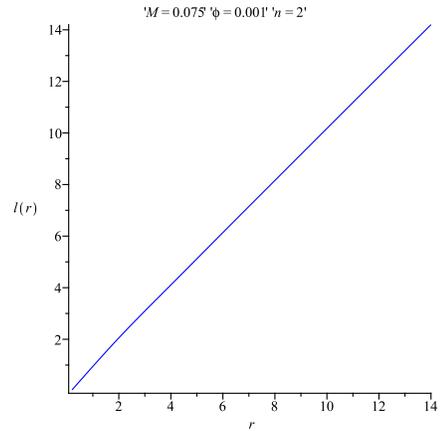}
       \caption{The radial proper length of the 4D wormhole has been shown against 'r'}
    \label{fig:1}
\end{figure}

Next our aim is to verify whether $b(r)$(see Fig.8) satisfies all the
physical requirement to form a shape function.
From $FIG.~9$ we see
that $\frac{b(r)}{r}\rightarrow 0 $ as $n\rightarrow\infty$ so
spacetime is asymptotically flat. The throat of the wormhole is the
point where $G(r)=b(r)-r$ cuts the $r$ axis.   From $FIG.~10 $ we
see that $G(r)$ cuts the 'r' axis at 0.16, so the throat of the
wormhole occurs at $r=0.16$ for 5D spacetime. Consequently,  we see that
$b'(0.16)=0.89<1$. For $r>r_0$, we see that $b(r)-r <0 $ which implies $\frac{b(r)}{r}<1$
for $r>r_0$. The slope   is still positive up to  $r_1 = 0.22 $ but soon it
becomes negative ( see fig. 11). So we have a valid wormhole solution from the throat up to a certain radius, say, $r_1$ =.22,  which  is a convenient cut-off for the wormhole. This indicates,  we get a restricted class of wormhole in five dimensional spacetime.
 ~In this case we can match our interior
solution to the exterior 5D line element given by
\[ds^{2}=-\left(1-\frac{2\mu}{r^2}\right)dt^{2}+\left(1-\frac{2\mu}{r^2}\right)^{-1}dr^{2}\]
\[+r^{2}(d\theta_1^{2}+\sin^{2}\theta_1d\theta_2^{2}+\sin^{2}\theta_1\sin^{2}\theta_2d\theta_3^{2})\]\\
where $\mu$, the mass of the 5D wormhole, is defined by $\mu=\frac{4Gm}{3\pi}$.

 To obtain the profile of embedding diagram and radial proper distance in $\theta_3= constant$ plane in  5D wormhole spacetime, we have to find out the integral of equations (17)and (18) respectively. But it is not possible to  perform it analytically. So we will take the help of numerical integration. The numerical values are
obtained by fixing a particular value of lower limit and by changing the upper limit, which are given in Table-III
 and Table-IV respectively. The embedding diagram $z(r)$ and the radial proper distance $l(r)$ of 5D wormhole have been depicted in fig.~13 and fig.~12 respectively. Fig.~14 represents the full visualization of 5D wormhole obtained by rotating the embedding curve  about 'z-axis'.

\subsection{Six and Seven dimensional Spacetime (n=4) and (n=5)}
In six dimensional spacetime,
the shape function of the wormhole is given by,
\begin{equation}
b(r)=\frac{1}{r^{2}}\left[\frac{M\sqrt{\phi}}{2\pi}\left\{\frac{3}{\sqrt{\phi}}\tan^{-1}\left(\frac{r}{\sqrt{\phi}}
 \right)-\frac{5r^{3}+3r\phi}{(r^{2}+\phi)^{2}}\right\}+C \right]
\end{equation}

The other parameters are given by
\begin{equation}
8\pi p_r=-\frac{6}{r^{5}}\left[\frac{M\sqrt{\phi}}{2\pi}\left\{\frac{3}{\sqrt{\phi}}\tan^{-1}
\left(\frac{r}{\sqrt{\phi}}  \right)-\frac{5r^{3}+3r\phi}{(r^{2}+\phi)^{2}}\right\}+C \right]
\end{equation}

\[8\pi p_t=\frac{3}{2r^{5}}\left[\frac{M\sqrt{\phi}}{2\pi}\left\{\frac{3}{\sqrt{\phi}}\tan^{-1}
\left(\frac{r}{\sqrt{\phi}}\right)-\frac{5r^{3}+3r\phi}{(r^{2}+\phi)^{2}}\right\}+C \right]\]
\begin{equation}
~~~~~~~~~~~~~~~~~~~~~~~~~~~~-\frac{6M\sqrt{\phi}}{\pi}\frac{1}{(r^{2}+\phi)^{3}}
\end{equation}

\begin{figure}[htbp]
    \centering
        \includegraphics[scale=.3]{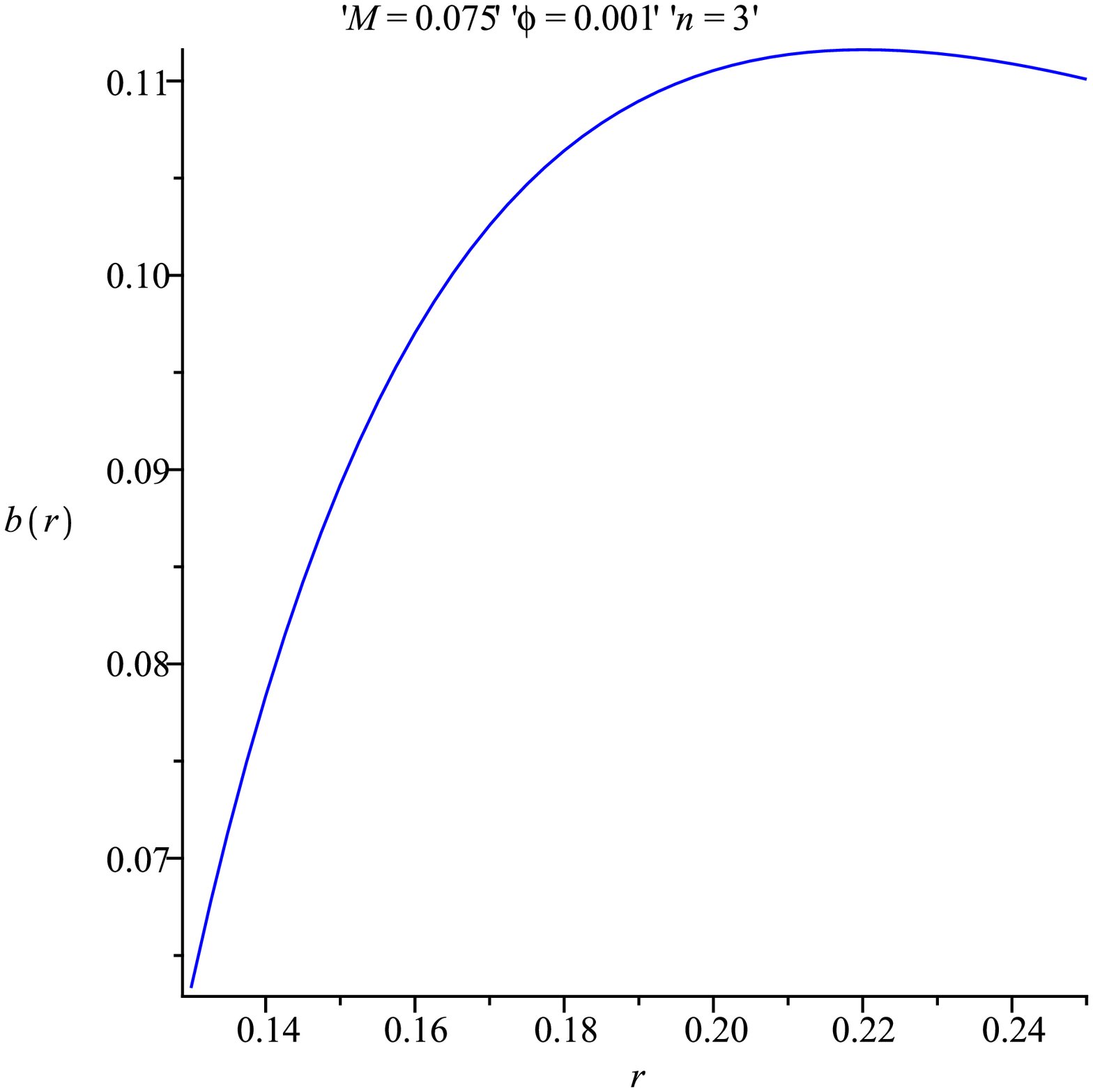}
       \caption{ Diagram of the shape function of the wormhole in five dimension is shown against $r$}
    \label{fig:1}
\end{figure}

\begin{figure}[htbp]
    \centering
        \includegraphics[scale=.3]{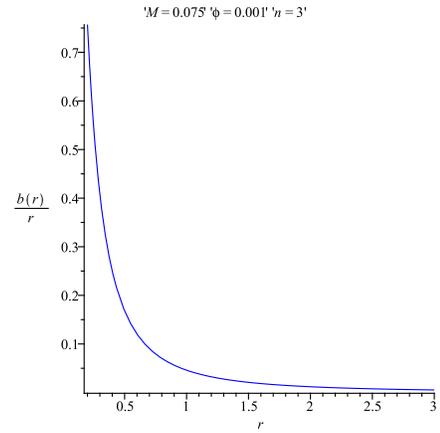}
       \caption{ Diagram of the flatness of the wormhole spacetime in five dimension is shown against $r$}
    \label{fig:1}
\end{figure}
\begin{table}
\caption{Values of $z(r)$ of 5D spacetime for different $r$. $ r_0^{+} =0.16 $,
$\phi=0.001$,$M=0.075$}

{\begin{tabular}{@{}cc@{}} \toprule $r$ & $l(r)$ \\
\colrule
0.17  &  ~~~~0.0221068505\\
0.175 &  ~~~~0.0388300822\\
0.18  &  ~~~~0.0527692279\\
0.19  &  ~~~~0.0758115765\\
0.20  &  ~~~~0.0948891894\\
0.21  &  ~~~~0.1114219209\\
\botrule
\end{tabular}}
\end{table}
\begin{table}
\caption{Values of $l(r)$ of 5D spacetime for different $r$. $ r_0^{+} =0.165 $,
$\phi=0.001$,$M=0.075$}

{\begin{tabular}{@{}cc@{}} \toprule $r$ & $l(r)$ \\
\colrule
0.17  &~~~~~~0.2267089806\\
0.18  &~~~~~~0.5493836396\\
0.19  &~~~~~~0.8006389468\\
0.20  &~~~~~~0.1016074351\\
0.21  &~~~~~~0.1209317737\\
\botrule
\end{tabular}}
\end{table}

\begin{figure}[htbp]
    \centering
        \includegraphics[scale=.3]{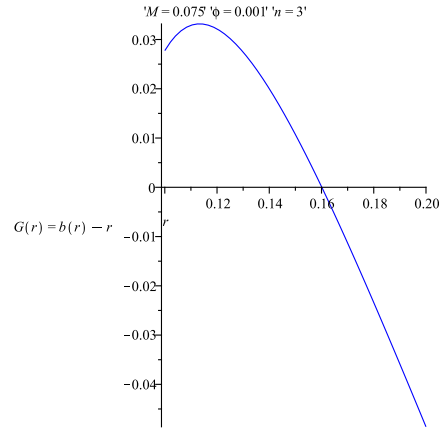}
       \caption{The throat of the wormhole in $5D$ case.}
    \label{fig:1}
\end{figure}

\begin{figure}[htbp]
    \centering
        \includegraphics[scale=.3]{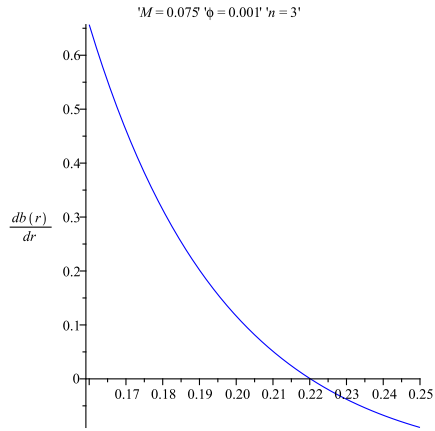}
       \caption{ The slope is still positive up to  $r_1 = 0.22 $ but soon
becomes negative.}
    \label{fig:1}
\end{figure}

\begin{figure}[htbp]
    \centering
        \includegraphics[scale=.3]{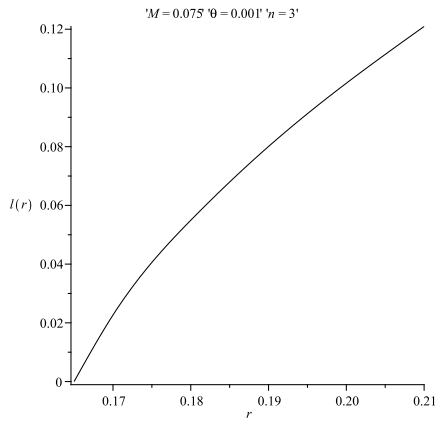}
       \caption{ The radial proper distance of the 5D wormhole in $\theta_3=0$ plane}
    \label{fig:1}
\end{figure}

Next we want to verify whether $b(r)$ obeys the physical conditions (see sec. III) to be a shape
 function of a wormhole. From $FIG.(16)$ we see that  $b(r)$ is a  strictly decreasing function of $r$, the
  same situation arises in seven dimensional spacetime $(n=5)$. In seven dimensional spacetime the shape function
   of the wormhole$(see FIG.~17)$ is given by,
\begin{equation}
b(r)=\frac{1}{r^{3}}\left[-\frac{16M\sqrt{\phi}} {75\pi}\frac{15r^{4}+20r^{2}\phi+8\phi^{2}}{(r^{2}+\phi)
^{\frac{5}{2}}} +C\right]
\end{equation}

\begin{figure}[htbp]
    \centering
        \includegraphics[scale=.3]{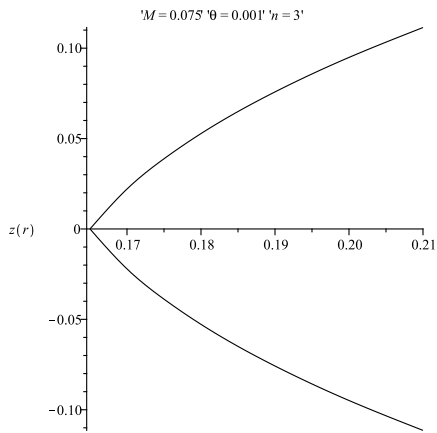}
       \caption{The embedding diagram of 5D wormhole in $\theta_3=0$ plane}
    \label{fig:1}
\end{figure}

\begin{figure}[htbp]
    \centering
        \includegraphics[scale=.3]{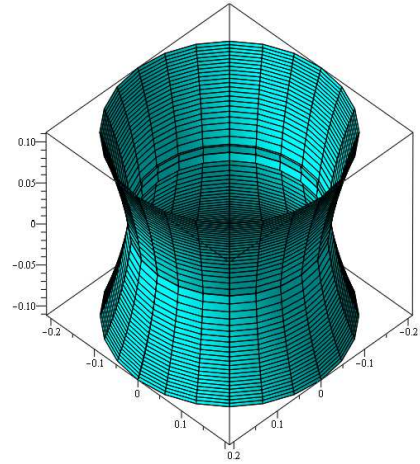}
       \caption{ The full visualization of 5D wormhole in $\theta_3=0$ plane}
    \label{fig:1}
\end{figure}

\begin{figure}[htbp]
    \centering
        \includegraphics[scale=.3]{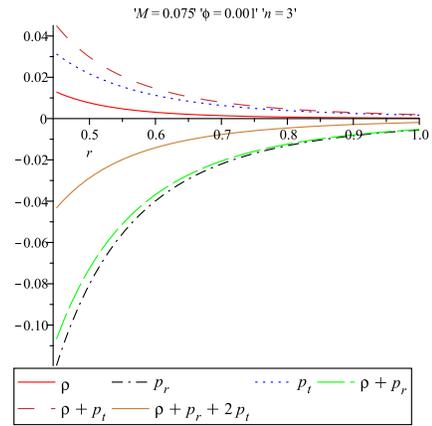}
       \caption{ Diagram of Energy condition for 5D wormhole has been plotted against $r$}
    \label{fig:1}
\end{figure}

\begin{figure}[htbp]
    \centering
        \includegraphics[scale=.3]{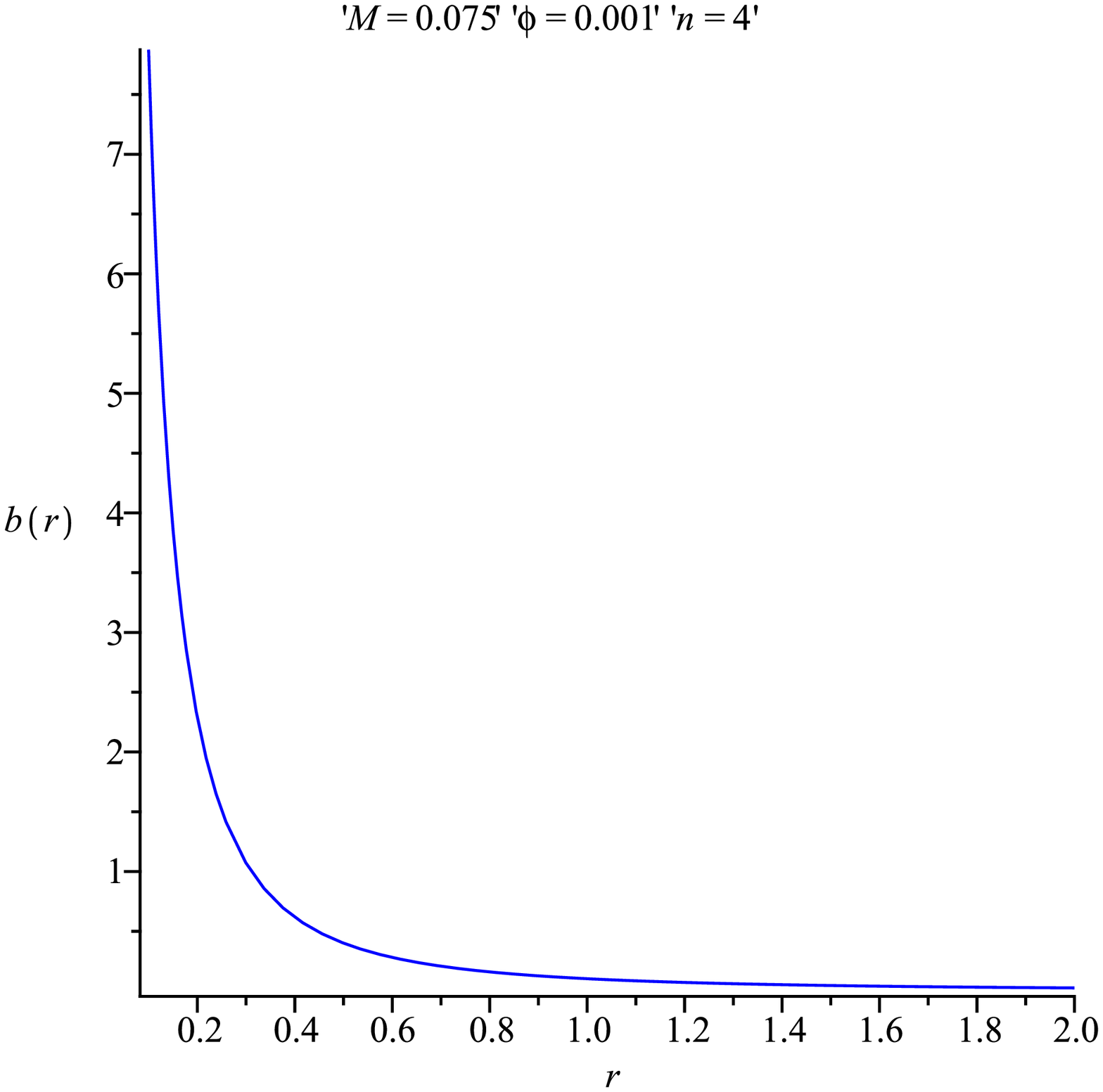}
       \caption{ Diagram of the shape function of the wormhole in six dimension is shown against $r$}
    \label{fig:1}
\end{figure}

\begin{figure}[htbp]
    \centering
        \includegraphics[scale=.3]{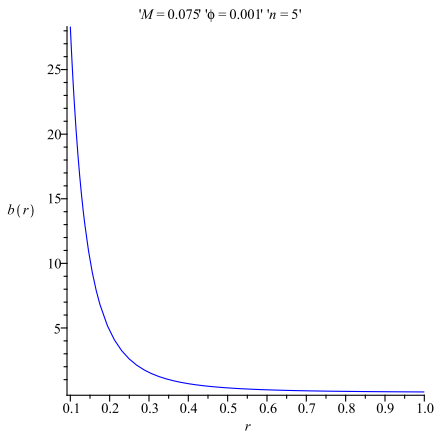}
       \caption{ Diagram of the shape function of the wormhole in seven dimension is shown against $r$}
    \label{fig:1}
\end{figure}

So it is clear that no wormhole solutions exist for $n>3$ because for $n>3$ the shape function $b(r)$~(see eq.(12)) contains a factor $\frac{1}{r^{n-2}}$ which is  a dominating one  and if we consider
  $n>3$ the shape function becomes monotonically decreasing, as a result $ b(r)$ in more than higher dimensional spacetime fails to be a shape
   function.

\begin{figure}[htbp]
    \centering
        \includegraphics[scale=.3]{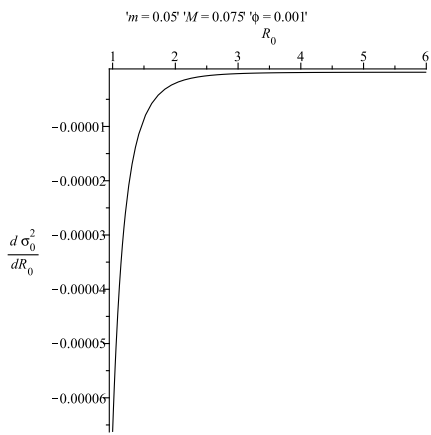}
       \caption{$\frac{d\sigma_0^{2}}{dR_0}$ has been plotted against $R_0$}
    \label{fig:1}
\end{figure}

\subsection{Energy Condition}
In this subsection,  we are going to verify whether our particular model   of wormholes(both 4D and 5D)  satisfies
all the energy conditions namely null energy condition (NEC), weak energy condition (WEC), strong energy condition
(SEC) stated as follows :
\begin{equation}
\rho\geq0
\end{equation}
\begin{equation}
\rho+p_r\geq0
\end{equation}
\begin{equation}
\rho+p_t\geq0
\end{equation}
\begin{equation}
\rho+p_r+2p_t\geq0
\end{equation}

The profile of energy conditions for four and five dimensional wormhole have been shown in $Fig.4$ and $Fig.15$
respectively. The figure indicates that $\rho+p_r <0$ i.e. the null energy condition(NEC) is violated   to hold a  wormhole open.\\

\begin{figure}[htbp]
    \centering
        \includegraphics[scale=.3]{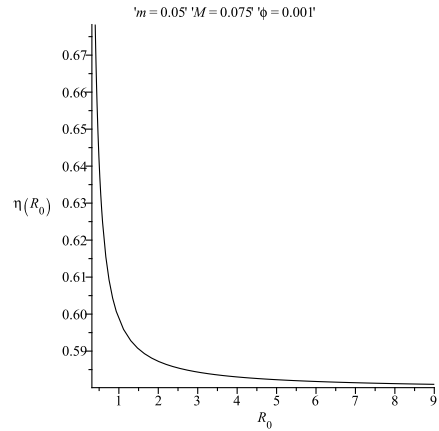}
       \caption{Plot of $ \eta(R_0)$ vs. $ R_0$}
    \label{fig:1}
\end{figure}

\section{Linearized stability analysis}

In this section, we will focus on the stability of the wormhole in four dimensional spacetime(n=2).

Using the expression of (20) in equation (22) and rearranging we get
\begin{equation}
\dot{R}^{2}+V(R)=0
\end{equation}
where $V(R)$ is given by
\[V(R)=F(R)-\left(\frac{m_s}{2R}\right)^{2}-\left(\frac{2m-b(R)}{2m_s(R)}\right)^{2}\]

where $F(R)=1-\frac{b+2m}{2R}$\\

({\bf for details calculation see the appendix.})\\

 ~~To discuss the linearized stability analysis let us take a linear perturbation around a static radius $R_0$.
Expanding $V(R)$ by Taylor series around the radius of the static solution $ R= R_0$,  one can obtain

\[V(R)=V(R_0)+(R-R_0)V'(R_0)+\frac{(R-R_0)^{2}}{2}V''(R_0)\]
\begin{equation}~~~~~~~~~~~~~~~~~~~~~~~~~~~+O[(R-R_0)^{3}]
\end{equation}
where 'prime' denotes derivative with respect to 'R' \\
Since we are linearizing around static radius $R=R_0$ we must have $V(R_0)=0,V'(R_0)=0$. The configuration will be
stable if $V(R)$ has a local minimum at $R_0$ i.e. if  $V''(R_0)>0$. \\
Now from the relation $ V'(R_0)=0$ we get,
\begin{widetext}
\begin{equation}
\left(\frac{m_s(R_0)}{2R_0}\right)'
=\left(\frac{R_0}{m_s(R_0)}\right)\left[F'(R_0)-2\left(\frac{2m-b(R_0)}{2m_s(R_0)}\right)\left(\frac{2m-b(R_0)}{2m_s(R_0)}\right)'\right]
\end{equation}
\end{widetext}

\[V''(R)=F''(R)-2\left(\frac{m_s(R)}{2R}\right)\left(\frac{m_s(R)}{2R}\right)''
-2\left\{\left(\frac{m_s(R)}{2R}\right)'\right\}^{2}\]
\[~~~-2\left(\frac{2m-b(R)}{2m_s(R)}\right)
\left(\frac{2m-b(R)}{2m_s(R)}\right)''-2\left\{\left(\frac{2m-b(R)}{2m_s(R)}\right)'\right\}^{2}\]
Now the configuration will be stable if $V''(R_0)>0$.\\
i.e.  if
\begin{equation}
\Omega_0>2\pi R_0 \sigma_0 \left(\frac{m_s(R)}{2R}\right)''|_{R=R_0}
\end{equation}
where $\Omega_0$ is given by
\[\Omega_0=\frac{F''(R_0)}{2}-\left\{\left(\frac{m_s(R_0)}{2R_0}\right)'\right\}^{2}
-\left(\frac{2m-b(R_0)}{2m_s(R_0)}\right)\times\]
\[\left(\frac{2m-b(R_0)}{2m_s(R_0)}\right)''-\left\{\left(\frac{2m-b(R_0)}{2m_s(R_0)}\right)'\right\}^{2}\]
Now using the expression of $\left(\frac{m_s}{2R_0}\right)''$ in equation $(40)$ we get,
\begin{equation}
\eta(R_0) \frac{d\sigma^{2}}{dR}|_{R=R_0}>\frac{1}{2\pi}\left[\sigma_0 \Xi(R_0) -\frac{\Omega_0}{2\pi R_0}\right]
\end{equation}
({\bf for details calculation see Appendix})\\
which gives,
\begin{equation}
\eta(R_0)>\Theta(R_0) \left(\frac{d\sigma^{2}}{dR}\right)^{-1}|_{R=R_0} ~~~~~~if ~~~~\frac{d\sigma^{2}}{dR}|_{R=R_0}>0
\end{equation}
and
\begin{equation}
\eta(R_0)<\Theta(R_0) \left(\frac{d\sigma^{2}}{dR}\right)^{-1}|_{R=R_0} ~~~~~~if ~~~~~~~\frac{d\sigma^{2}}{dR}|_{R=R_0}<0
\end{equation}
 ~~~where $ \Theta(R_0)=\frac{1}{2\pi}\left[\sigma(R_0) \Xi(R_0) -\frac{\Omega_0}{2\pi R_0}\right]$\\

 ~~For   wormhole in four dimensional spacetime,  we have $\frac{d\sigma_0^{2}}{dR_0}<0$ (see fig.$18 $). Therefore, we conclude that   the
stability of the wormhole is given by equation $(43)$. \\
  Next we are interested to search the region where our 4D wormhole model is stable. To obtain the region we have plotted the graph of $\eta(R_0)=\Theta(R_0) \left(\frac{d\sigma^{2}}{dR}\right)^{-1}|_{R=R_0}$ which has been shown in $Fig.~19$. Since the stability region is given by the inequality $\eta(R_0)<\Theta(R_0) \left(\frac{d\sigma^{2}}{dR}\right)^{-1}|_{R=R_0}$. So the region of stability is given below the curve. Interestingly, we see that $0<\eta\le1$ which indicates that our wormhole is very much stable.

\section{Active Mass Function}
We will discuss the active gravitational mass within the region
from the throat $r_0$ up to the radius $R$ for two cases because
we have seen earlier that wormholes exist  only in  four and five
dimensional spacetimes.  For 4D case the active gravitational
mass can be obtained as,
\[M_{active}=\int_{r_0}^{r}4\pi \rho r^{2}\]
\begin{equation}~=\frac{2M}{\pi}\left[tan^{-1}\left(\frac{r}{\sqrt{\phi}}\right)
-\frac{r\sqrt{\phi}}{r^{2}+\phi}\right]_{r_0}^{r}\end{equation}

For 5D case, one can get  the active gravitational mass as,
\begin{equation}M_{active}=\int_{r_0}^{r}2\pi^{2}r^{3}\rho dr=2M\sqrt{\phi}
\left[-\frac{1}{3}\frac{3r^{2}+2\phi}{(r^{2}+\phi)^{\frac{3}{2}}}\right]_{r_0}^{r}\end{equation}
\begin{figure}
        \includegraphics[scale=.30]{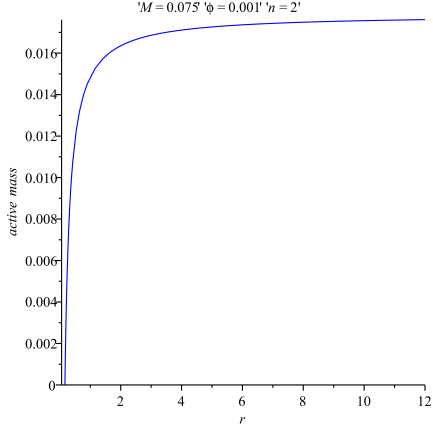}
        \caption{The variation of $M_{active}$ for 4D case is shown against $r$.}
   \label{fig:wh20}
\end{figure}
\begin{figure}
        \includegraphics[scale=.30]{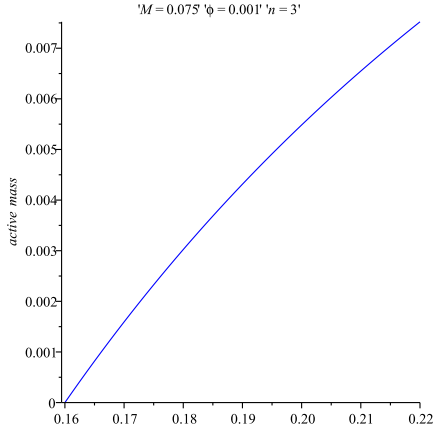}
        \caption{The variation of $M_{active}$  for 4D is shown against $r$ in the restricted range.}
   \label{fig:wh20}
\end{figure}

To see the nature of the active gravitational mass $M_{active}$
of the wormhole, we have drawn the figures 20 and 21 for four and
five dimensional spacetimes respectively. The positive nature
indicates that the models are physically acceptable.

\section{Total Gravitational Energy}

Following  Lyndell-Bell et al. \cite{Lyndell2007} and Nandi et al.
\cite{Nandi2009} prescription,  we are trying  to calculate the
total gravitational energy of the wormholes for four and five
dimensional spacetimes. For 4D case we have,
\begin{equation}E_g=\frac{1}{2}\int_{r_0}^{r}[1-(g_{rr})^{\frac{1}{2}}]\rho r^{2} dr
 +\frac{r_0}{2}.\end{equation}

And for 5D case we have,
\begin{equation}E_g=\frac{\pi}{4}\int_{r_0}^{r}[1-(g_{rr})^{\frac{1}{2}}]\rho r^{3}
 dr +\frac{ \pi r_0}{4}, \end{equation}
where
\[g_{rr}=\left(1-\frac{b(r)}{r}\right)^{-1}\]
  Since, it not possible to find analytical solutions of thgese integrals,
  we solve it numerically assuming  the range of the integration
  from the throat $r_0$ to the embedded radial
space of the wormhole geometry (see tables 5 and 6). Figures (22)
and (23) show that $E_g > 0$ which indicate   that there is a
repulsion around the throat. This nature of $E_g $    is expected
for construction of a physically valid wormhole.

\begin{table}
\caption{Values of $E_g$ of 4D wormhole for different $r$. $ r_0
=0.165 $, $\phi=0.001$,$M=0.075$}

{\begin{tabular}{@{}cc@{}} \toprule $r$ & $E_g$ \\
\colrule
0.2 &  ~~~~0.08248549261\\
0.3  &  ~~~~0.08241864488\\
0.4  &  ~~~~0.08234761560\\
0.5  &  ~~~~0.08227653015\\
0.6 &  ~~~~0.08220571804\\
0.7 &  ~~~~0.08213515524\\
0.8 &  ~~~~0.08206478498\\
0.9 &~~~~~~0.08199455998\\
1.0  &~~~~~0.08192444534\\
\botrule
\end{tabular}}
\end{table}

\begin{table}
\caption{Values of $E_g$ of 5D wormhole for different $r$. $ r_0
=0.165 $, $\phi=0.001$,$M=0.075$}

{\begin{tabular}{@{}cc@{}} \toprule $r$ & $E_g$ \\
\colrule
0.17 &  ~~~~0.1255974565\\
0.175  &  ~~~~.1255950236\\
0.18  &  ~~~~0.1255922125\\
0.185  &  ~~~~.1255891335\\
0.19 &  ~~~~0.1255858558\\
0.195 &  ~~~~0.1255824265\\
0.20 &  ~~~~.1255788793\\
0.205 &~~~~~~0.1255752392\\
0.21  &~~~~~0.1255715251\\
\botrule
\end{tabular}}
\end{table}

\begin{figure}
\centering
\includegraphics[scale=.30]{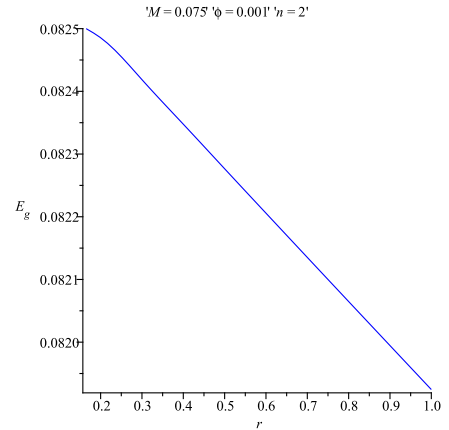}
\caption{The variation of $E_g$ for 4D case is shown against $r$.}
\label{fig:E_g}
\end{figure}
\begin{figure}
\centering
\includegraphics[scale=.30]{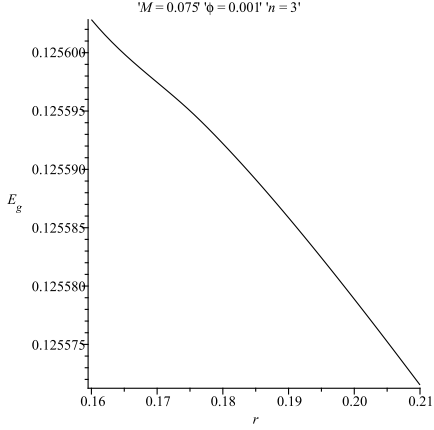}
\caption{The variation of $E_g$ for 5D case  is shown against $r$
in the restricted range.} \label{fig:E_g}
\end{figure}

\section{Discussion and Concluding remarks}
In the present paper,  we have obtained a new class of wormhole solutions in the context of noncommutative geometry
background. In this paper we have chosen Lorentzian distribution
  function as the density function of the wormhole in noncommutative inspired spacetime. We have examined whether    wormholes
   with Lorentzian distribution exist in different  dimensional spacetimes. From the above investigations,  we see that the wormholes exist  only in   four and
   five dimensional spacetimes. In case of five dimension,  we have observed that wormhole exists in
   a very restricted region.  From six dimension and onwards the shape function of the wormhole
    becomes monotonic decreasing due to the presence of  the term $\frac{1}{r^{n-2}}$ in the shape function. So from the above
     discussion,  we can conclude that no wormhole solution exists beyond five  dimensional spacetime. In case of four and five
      dimensional wormholes $\rho+p_r<0$ i.e.  the null energy conditions are  violated. We note that four dimensional wormhole is large enough, however,   one can  get five dimensional wormhole geometry only in a very restricted region.
      We have match our interior wormhole solution to the exterior schwarzschild
       spacetime in presence of thin shell. The linearized stability analysis under small radial perturbation has also
        been discussed for four dimensional wormhole.
         We have provided the region where the wormhole is stable and with the
          help of graphical representation we have proved that
           $0<\eta\le1$ i.e. our wormhole is  very much stable.
           Finally, we can   compare our results with those obtained in
           reference \cite{farook5}.
           In both models ( our and in reference \cite{farook5} ),
            noncommutativity replaces point-like structures by smeared objects.
             In reference \cite{farook5},
               they had  used  the Gaussian distribution as the density function
                in the noncommutative inspired  spacetime , whereas,  we have used
                Lorentzian distribution as the density function in the noncommutative
                 inspired  spacetime.
In both the cases,  it is  shown that wormhole solutions exist
only four and five dimensional spacetimes.  However, our study is
more complete compare to the reference \cite{farook5}. We have
studied stability and calculated active gravitational mass and
total gravitational energy of the wormhole. Thus, we can comment
that that this study confirms the results that that wormhole
solutions exist in non commutative inspired geometry only for
four and five dimensional spacetimes.

\section{Acknowledgement}

FR is thankful to the Inter University Centre for Astronomy and As
trophysics (IUCAA), India, for providing research facilities. We are thankful to the referee for his valuable suggestions.\\

{\bf Appendix:1}
\[m_s=4\pi R^{2}\sigma\]
using the expression of $\sigma$ we get,
\[or~~\frac{m_s}{4\pi R^{2}}=\frac{1}{4\pi R}\left[\sqrt{1-\frac{b(R)}{R}+\dot{R}^{2}}
-\sqrt{1-\frac{2m}{R}+\dot{R}^{2}}\right]\]
\[or,~~\frac{m_s}{R}=\sqrt{1-\frac{b(R)}{R}+\dot{R}^{2}}-\sqrt{1-\frac{2m}{R}+\dot{R}^{2}}\]
\[or~~~\frac{m_s}{R}-\sqrt{1-\frac{b(R)}{R}+\dot{R}^{2}}=-\sqrt{1-\frac{2m}{R}+\dot{R}^{2}}\]
Squaring bothside we get,
\[\left(\frac{m_s}{R}\right)^{2}-2\frac{m_s}{R}\sqrt{1-\frac{b(R)}{R}+\dot{R}^{2}}=\frac{1}{R}(b-2m)\]
\[or,~~~~~~~\frac{m_s}{R}\left[\frac{m_s}{R}-2\sqrt{1-\frac{b(R)}{R}+\dot{R}^{2}}\right]=\frac{1}{R}(b-2m)\]
\[or,~~~\frac{m_s}{R}-2\sqrt{1-\frac{b}{R}+\dot{R}^{2}}=\frac{1}{m_s}(b-2m)\]
\[or,~~~\frac{m_s}{2R}-\frac{b(R)-2m}{2m_s}=\sqrt{1-\frac{b(R)}{R}+\dot{R}^{2}}\]\\

~~~~~~again squaring bothside we get,

\[\dot{R}^{2}=\left(\frac{m_s}{2R}\right)^{2}+\left(\frac{b(R)-2m}{2m_s}\right)^{2}+\frac{b+2m}{2R}-1\]
Now $\dot{R}^{2}=-V(R)$
which gives,
\[V(R)=1-\frac{b+2m}{2R}-\left(\frac{m_s}{2R}\right)^{2}-\left(\frac{b-2m}{2m_s}\right)^{2}\]
\[V(R)=F(R)-\left(\frac{m_s}{2R}\right)^{2}-\left(\frac{b-2m}{2m_s}\right)^{2}\]
where $$F(R)=1-\frac{b+2m}{2R}$$\\

{\bf Appendix:2}

\[m_s=4\pi R^{2}\sigma \]
\[or,~~\frac{m_s}{2R}=2\pi R \sigma\]
Differentiating bothside with respect to R we get,
\[or,~~\left(\frac{m_s}{2R}\right)'=2\pi (R \sigma'+\sigma)\]
\[~~~~~~~~~~~~~~~~~~~~~~~~~~~~~~~~~~~~~~=2\pi R\left\{-\frac{2}{R}(\sigma+\mathcal{P})+\Xi \right\}+2\pi \sigma\]
\[~~~~~~~~~~~~~~~~~~~=-4\pi\mathcal{P}+2\pi R \Xi-2\pi\sigma\]
Differentiating bothside with respect to R we get,
\[\left(\frac{m_s}{2R}\right)''=-4\pi \mathcal{P}'+2\pi (R \Xi'+\Xi)-2\pi\sigma'\]
Using the value of $\sigma'$ we get,
\[\left(\frac{m_s}{2R}\right)''=-4\pi \mathcal{P}'+2\pi (R \Xi'+\Xi)-2\pi\left\{-\frac{2}{R}(\sigma+\mathcal{P})+\Xi
 \right\}\]
\[=\frac{4\pi}{R}(\sigma+\mathcal{P})+2\pi R\Xi'-4\pi\eta\sigma'\]
therefore,
\[\left(\frac{m_s}{2R}\right)''=\Upsilon-4\pi \eta\sigma'\]
where,
\[\Upsilon=\frac{4\pi}{R}(\sigma+\mathcal{P})+2\pi R\Xi'\]

{\bf Appendix:3}

\[V''(R)=F''(R)-2\left(\frac{m_s(R)}{2R}\right)\left(\frac{m_s(R)}{2R}\right)''
+2\left\{\left(\frac{m_s(R)}{2R}\right)'\right\}^{2}\]
\[~~~~~~~~-2\left(\frac{2m-b(R)}{2m_s(R)}\right)
\left(\frac{2m-b(R)}{2m_s(R)}\right)''+2\left\{\left(\frac{2m-b(R)}{2m_s(R)}\right)'\right\}^{2}\]
Now $V''(R_0)>0$ gives

\[\Omega_0>2\pi R_0\sigma_0 \left(\frac{m_s}{2R_0}\right)''\]
Where \[\Omega_0=\frac{F''(R_0)}{2}-2\left\{\left(\frac{m_s(R_0)}{2R_0}\right)'\right\}^{2}-\left(\frac{2m-b(R_0)}
{2m_s(R_0)}\right)\times\]
\[\left(\frac{2m-b(R_0)}{2m_s(R_0)}\right)''
-\left\{\left(\frac{2m-b(R_0)}{2m_s(R_0)}\right)'\right\}^{2}\]
\[\Omega_0>2\pi R_0\sigma[\Xi(R_0)-4\pi\eta(R_0)\sigma_0']\]
\[\Omega_0>2\pi R_0\sigma_0\Xi(R_0)-4\pi^{2}R_0\eta(R_0) \frac{d\sigma_0^{2}}{dR_0}\]
\[\frac{\Omega_0}{2\pi R_0}>\sigma(R_0)\Xi(R_0)-2\pi\eta(R_0)\frac{d\sigma_0^{2}}{dR_0}\]
\[2\pi\eta(R_0)\frac{d\sigma_0^{2}}{dR_0}>\sigma_0\Xi(R_0)-\frac{\Omega_0}{2\pi R_0}\]
which gives,
\[\eta(R_0)\frac{d\sigma_0^{2}}{dR_0}>\frac{1}{2\pi}\left[\sigma_0\Xi(R_0)-\frac{\Omega_0}{2\pi R_0}\right]\]

\end{document}